\documentclass{article}
\usepackage{titlesec}
\usepackage[utf8]{inputenc}
\usepackage{graphicx} 
\usepackage{tikz}
\usepackage{amsmath}
\usepackage{amssymb}
\usepackage{authblk}
\usepackage{biblatex}
\usepackage[colorlinks=true, urlcolor=blue, linkcolor=blue, citecolor=red]{hyperref}
\title{Imperfect Graphs from Unitary Matrices - I}
\author[1]{Wesley Lewis}
\author[1]{Darsh Pareek}
\author[1]{Umesh Kumar}

\author[1,2]{Ravi Janjam}
\affil[1]{Researcher}
\affil[2]{Principal Investigator / Chief of R\&D}
\affil[1]{\href{https://www.numerikal-labs.org/}{Numerikal Labs}, 1942 Broadway St, Suite 314C, Boulder, CO 80302}
\begin{document}
\maketitle
\begin{abstract}
Matrix representations of quantum operators are computationally complete but often obscure the structural topology of information flow within a quantum circuit \cite{nielsen2000}. In this paper, we introduce a generalized graph-theoretic framework for analyzing quantum operators by mapping unitary matrices to directed graphs; we term these structures \emph{Imperfect Graphs} or more formally as \emph{Topological Structure of Superpositions}(TSS) as a tool to devise better Quantum Algorithms.\\

\indent In this framework, we represent computational basis states as vertices. A directed edge exists between two vertices if and only if there is a non-zero amplitude transition between them, effectively mapping the support of the unitary operator. In this paper we deliberately discard probability amplitudes and phase information to isolate the connectivity and reachability properties of the operator.\\

\indent We demonstrate how TSS intuitively helps describe gates such as the Hadamard, Pauli-(X,Y,Z) gates, etc \cite{nielsen2000}. This framework provides a novel perspective for viewing quantum circuits as discrete dynamical systems \cite{childs2009,aharonov2001}. \\

\end{abstract}

\newpage

\section{Introduction}
Quantum computing is fundamentally described by linear algebra, where quantum states are vectors in a Hilbert space and operations also referred to as Quantum Gates are represented by unitary matrices \cite{nielsen2000}. While this matrix formulation is essential for simulation, it lacks intuitive transparency regarding the structural behavior of a circuit. As the number of qubits $n$ increases, the Hilbert space grows to $2^n$ for two state realizations of qubits, making the interpretation of unitary matrices through visual inspection alone increasingly difficult.\\

To address this complexity, we can categorize the analysis of quantum operations into two primary approaches. First, the Circuit Elements Approach (CEA) constructs the system via independent, local operators (gates) added sequentially. While amenable to local modifications and hardware implementations, CEA often obscures the global topological properties of the final transformation when all state combinatorics are to be taken into account. Second, the Unitary Matrix Approach (UMA) analyzes the properties of the final unitary matrix $U$ encompassing the entire circuit. While UMA "natively" describes the global computational structure, extracting meaningful patterns from these high-dimensional matrices remains a challenge. 
For instance, the Solovay-Kitaev theorem \cite{dawson2006} provides a widely used algorithm for approximating a unitary matrix into smaller circuit elements \cite{shende2006,vatan2004}. However, while effective for compilation, these decomposition methods do not inherently reveal the algorithmic substructures or the topological information flow that a practitioner seeks to understand.
In classical computing, state-transition diagrams and control flow graphs are standard tools. These visualizations allow researchers to identify loops, dead ends, and connectivity properties at a glance. In the quantum domain however, constructing such representations is non-trivial, as a single basis state can evolve into a linear combination of all possible states with complex-valued amplitudes \cite{nielsen2000}. \\

In this paper, we propose topological abstraction of quantum circuits that leverages the Unitary Matrix Approach where all possible state transition combinatorics is considered. We introduce a framework to map unitary operators to directed graphs, which we term the Topological Structure of Superpositions (TSS). By treating the non-zero support of a unitary matrix as an adjacency matrix for a directed graph, we analyze quantum gates as discrete dynamical systems \cite{childs2009,aharonov2001}, isolating the connectivity properties of the operator from its probabilistic factors.

\section{Methodology}

\newcommand*{\CC}{%
    \mathbb{C}%
}
\newcommand*{\ZZ}{%
    \mathbb{Z}%
}
\newcommand{\ket}[1]{\ensuremath{\left|#1\right\rangle}}
\newcommand{\bra}[1]{\ensuremath{\left\langle#1\right|}}
\subsection{Theoretical Basis}
We consider a quantum system described by a Hilbert Space $\mathcal{H}$ of dimension $N = 2^n$ (for $n$ qubits) \cite{nielsen2000}. The computational basis $\mathcal{B} = \{e_0, e_1, \dots, e_{N-1}\}$ forms an orthonormal set spanning $\mathcal{H}$. A general quantum state vector $\ket{\psi}$ is defined as a linear combination of these basis vectors:
$$
\ket{\psi} = \sum_{i = 0}^{N-1}c_i e_i
$$
where $c_i \in \CC$ are complex probability amplitudes satisfying $\sum |c_i|^2 = 1$ \cite{nielsen2000}.
Quantum gates are represented by unitary matrices $U$ of size $N \times N$ that transform an input state to an output state: $\ket{\psi_{out}} = U\ket{\psi_{in}}$ \cite{nielsen2000}. While the final state vector $\ket{\psi_{out}}$ is mathematically defined by orthogonality, its physical interpretation is often multifaceted. The distribution of non-zero basis states in the output allows for various structural interpretations, including entanglement \cite{hein2004}, where the system admits multiple representations for the same quantum property. Our topological approach aims to visualize these representations explicitly.
\subsection{Graph Representation}
To analyze the topological structure of these operators, we construct a directed graph that maps the non-zero support of the unitary matrix. Unlike standard state-space visualizations which often require continuous parameters \cite{vandennest2004,raussendorf2001}, our mapping is discrete and topological. We define a mapping $\phi$ that associates each basis vector $e_i$ with a unique integer vertex index:
$$
\phi(e_i) = i \quad \text{for } i \in \{0, \dots, N-1\}
$$
This strictly limits our vertices to fundamental basis states. We deliberately exclude superposed states (linear combinations where multiple $c_i \neq 0$) from the vertex set to avoid the combinatorial explosion associated with power-set constructions.
\subsection{TSS Construction}
We construct the \textbf{Topological Structure of Superpositions (TSS)}, denoted as $G_{\text{TSS}} = (V, E)$, as follows:
\begin{itemize}
\item \textbf{Vertices ($V$):} The set of integers representing the computational basis states.
$$
V = \{0, 1, \dots, N-1\}
$$
The graph size corresponds linearly to the Hilbert space dimension $N$.
\item \textbf{Edges ($E$):} A directed edge $(j, i)$ exists from vertex $j$ to vertex $i$ if the unitary operator $U$ facilitates a transition from basis state $\ket{j}$ to basis state $\ket{i}$ with non-zero amplitude.
$$
E = \{ (j, i) \mid |\bra{i} U \ket{j}| > 0 \}
$$
\end{itemize}
\section{Statistical Analysis of TSS}
In this section, we present the topological analysis of TSS graphs generated by various fundamental unitary operators. Our method is universally applicable to any unitary matrix. We observed that the topological "shape" of the graph correlates strongly with the operator's role in quantum algorithms \cite{childs2009,aharonov2001}.\\

One should also understand that TSS is constructed based on the Gate's operation on the operand only once and on a single qubit. Analyzing the outcome on \emph{entangled vs superposition} states would lead to a different structure completely. So, this TSS is the \emph{fundamental variant of TSS} for anything that's expected to come in the future. \\

\subsection{Sparsity vs. Connectivity}
A key observation from our analysis is the trade-off between graph sparsity and connectivity. We choose two unique and widely studied matrices for demonstrating the idea. While the matrices themselves don't necessarily imply much, they however have some properties that are global in scope. \\
\begin{itemize}
    \item \textbf{Algorithmic Sparsity:} We observe that operators used for logical manipulation (such as control sequences or arithmetic functions) typically exhibit \textbf{sparse TSS connections}. For devising complex algorithms, high connectivity is often undesirable; if a single state vector immediately transitions into a superposition of all possible states, the system loses the structural specificity required for logical operations \cite{gottesman1997,aaronson2004}.
    \item \textbf{Data Loading Efficiency:} Conversely, highly connected graphs are advantageous for "Data Loading" phases. If a single input vector introduces transitions to a vast number of states (high out-degree), tangled controlled-gate operations can be utilized to load data efficiently into the system.
\end{itemize}

\subsection{Hadamard Matrices: The Normalizer}
The Hadamard operator is topologically distinct due to its density \cite{nielsen2000}. Since Hadamard matrices are constructed primarily of $\pm 1$ entries (up to a normalization factor), the operation on a basis state results in a final vector with non-zero entries at nearly every index.
\begin{itemize}
    \item \textbf{Topology:} The TSS of a Hadamard operator approaches a \textbf{fully connected graph} (or clique). Every node has a directed edge to almost every other node.
    \item \textbf{Interpretation:} The Hadamard matrix introduces the property of \textbf{normalization} to the TSS of any graph. It maximizes the topological entropy, ensuring the state is distributed across the entire Hilbert space.
\end{itemize}
\subsection{Pauli Matrices: Island Graphs}
Pauli gates ($X, Y, Z$) and their tensor products provide the most structured TSS topologies \cite{nielsen2000}.
\begin{itemize}
    \item \textbf{Topology:} The TSS per state for Pauli matrices generally manifests as \textbf{"forks."} Mathematically, this corresponds to sparsely connected between any two edges edges ($i \to j$ and $j \to i$) or disjoint cycles of length 2.
    \item \textbf{Interpretation:} This structure confirms the Pauli group's role as reversible, classical-like permutations \cite{gottesman1997}. Any tensor product of other matrices with a set of Pauli gates will inherit this structure  with, sparse topology. 
\end{itemize}

\subsection{Interpretation of Entanglement}
While the orthogonality of the final state vector is the primary mathematical constraint \cite{nielsen2000}, the TSS allows for broader physical interpretations. The specific connectivity pattern—where a single node branches into a specific subset of nodes—can be interpreted as a structural representation of \textbf{entanglement} \cite{hein2004,gross2009}. The graph topology reveals how the system maintains multiple representations (or paths) for the same quantum property, a feature not visible in the raw matrix values alone.

\subsection{TSS Examples}
In this section, we'll discuss some of the TSS graphs produced from widely used circuits and also some matrices that don't yet have a circuit representation or rather purely abstract in nature. \\

We made a deliberate choice of the unitary matrices for this paper to demonstrate how TSS changes. Complete TSS of fully connected graph of just $3$ or $4$ qubits is sufficiently complicated that we need a completely new visualization technique, and \emph{measures} for analysis. So, we only show a  few samples in this paper. \\

\noindent
$\mathbf{P_x\otimes P_x\otimes P_x \otimes P_x}$\label{sec:pxpxpxpx} \\

We get a matrix whose dimensions are $2^4\times 2^4$ with the highest state vector represented in quantum state notation $\ket{1111} \sim \ket{15}$ \\

\begin{figure}[h]
    \centering
    \includegraphics[width=1\linewidth]{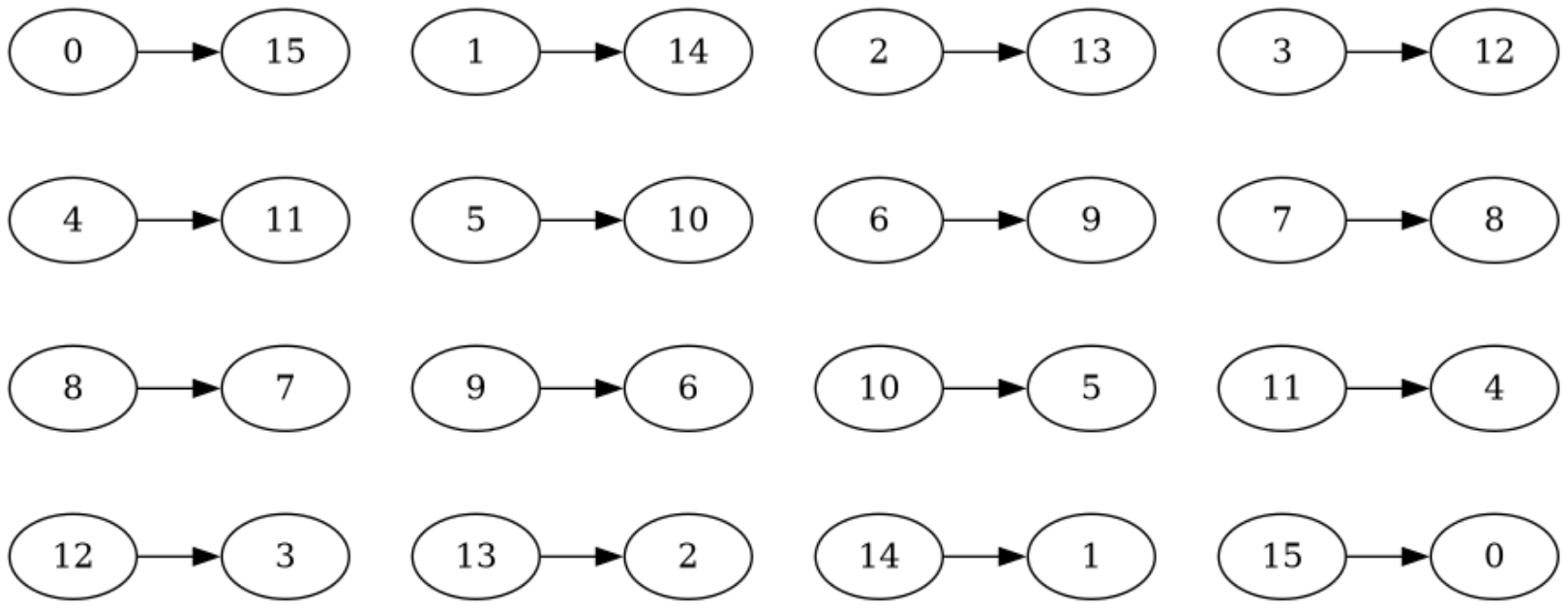}
    \caption{TSS Graph $\bigotimes_{i=1}^4$Pauli\_X }
    \label{fig:pxpxpxpx}
\end{figure}

$P_x$ refers to the Pauli-X matrix. This construction is an arbitrary choice without a circuit representation and no known functional utility either. The TSS graph [\ref{fig:pxpxpxpx}] provides a rich structure with pairs of nodes connected in both ways. It is also noticeable that, the sum of the states equals $15$ no matter which pair we choose, where each state has some complementary state that completes the sum to $15$. This property can be used for some purpose. \\

\noindent
$\mathbf{\mathrm{\textbf{Berkeley}}\otimes P_x}$\\
We get a matrix whose dimensions are $2^3\times 2^3$ with the highest state vector represented in quantum state notation $\ket{111} \sim \ket{7}$. \\

\begin{figure}[h]
    \centering
    \includegraphics[width=1\linewidth]{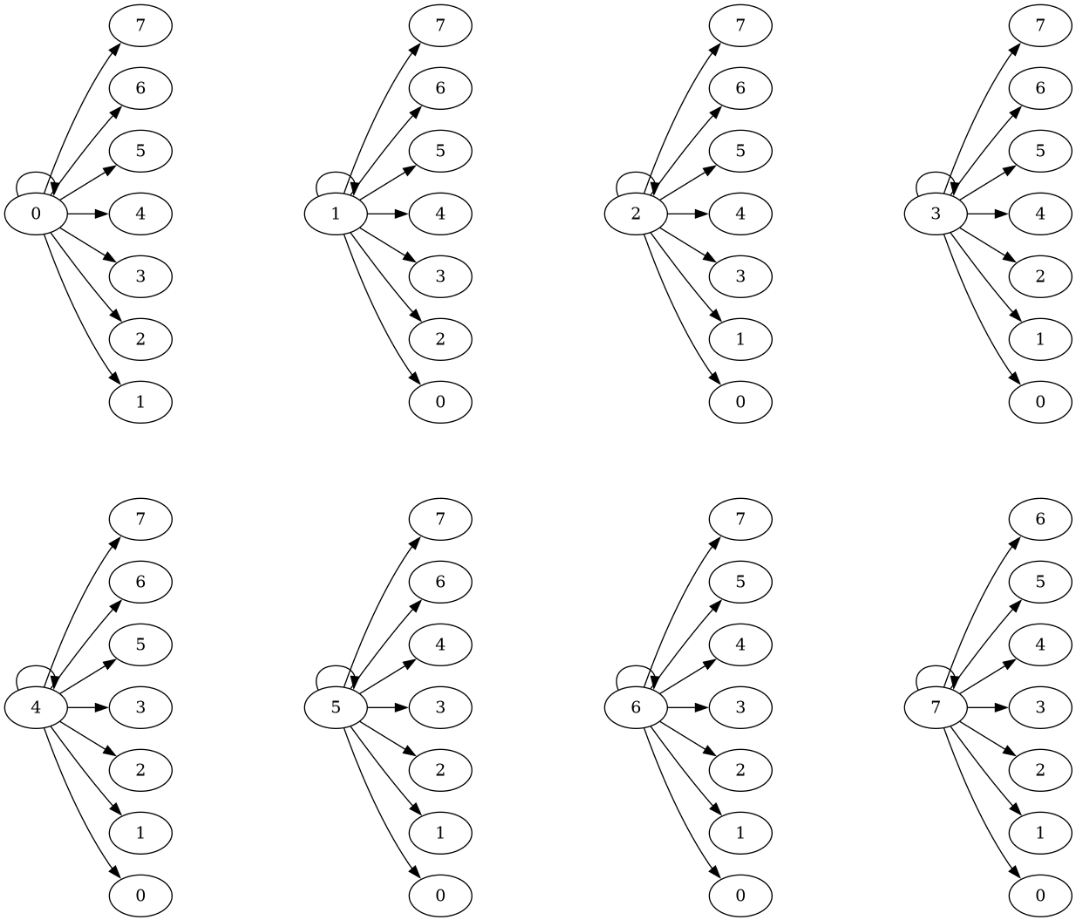}
    \caption{Node Level TSS: Berkeley $\otimes\; P_x$}
    \label{fig:ber_px}
\end{figure}

TSS structure [\ref{fig:ber_px}] for all the nodes are isomorphic to each other, and it seems more like that of the TSS of a simple Hadamard matrix. It's worth noting that $P_x$ didn't introduce any sparsity. \\

\noindent
$\mathbf{\mathrm{\textbf{Berkeley}}\otimes \mathrm{\textbf{Berkeley}}}$\label{sec:bb}\\
We get a matrix whose dimensions are $2^4\times 2^4$ with the highest state vector represented in quantum state notation $\ket{1111} \sim \ket{15}$. \\

\begin{figure}[h]
    \centering
    \includegraphics[width=1\linewidth]{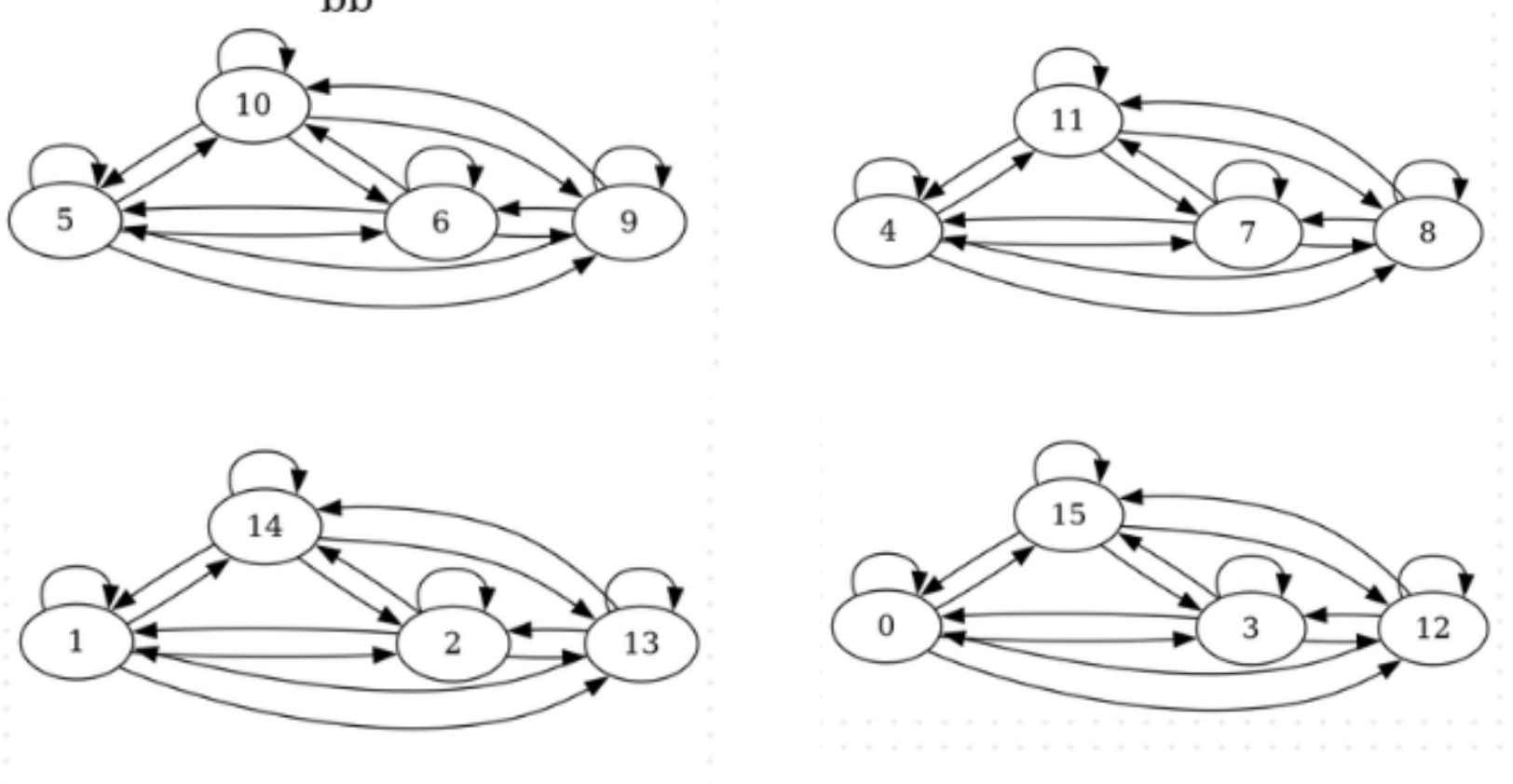}
    \caption{TSS Graph: Berkeley $\otimes$ Berkeley}
    \label{fig:bb_full}
\end{figure}

\begin{figure}[h]
    \centering
    \includegraphics[width=1\linewidth]{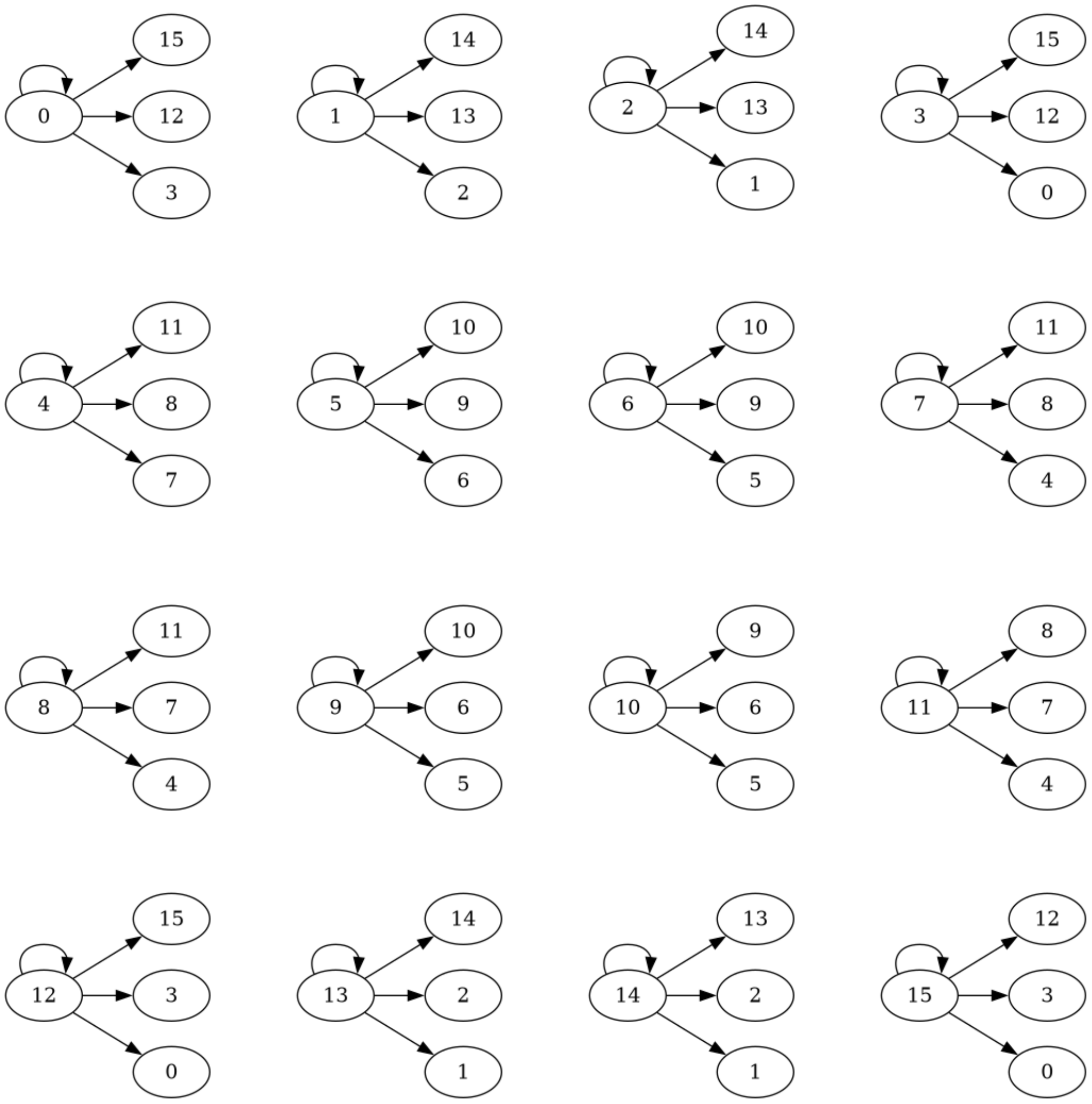}
    \caption{Node Level TSS: Berkeley $\otimes$ Berkeley}
    \label{fig:bb_node}
\end{figure}

TSS structure [\ref{fig:bb_full},\ref{fig:bb_node}] for all nodes are isomorphic to each other where we have quadruples in the outputs. The reason for such a symmetric structure is easy to understand. Berkeley matrix is symmetric in shape with not many unique elements which implies, any vector multiplied to this gate will lead to unit values showing up in the state vector in positions that are symmetric as well. \\ 

Berkeley matrix has a structure as given below with two unique values complex conjugated. When a multiplication with a state vector happens, it's easily seen how things show up depending on the position where the unit in the decimal notation of qubit is located. 
$$
\begin{bmatrix}
a &0 &0 &b \\
0 &c &d &0 \\
0 &e &f &0 \\
g &0 &0 &h \\
\end{bmatrix}
$$

\noindent
$\mathbf{\mathrm{\textbf{Grover}}\otimes P_x}$

We get a matrix whose dimensions are $2^3\times 2^3$ with the highest state vector represented in quantum state notation $\ket{111} \sim \ket{7}$. \\

\begin{figure}[h]
    \centering
    \includegraphics[width=1\linewidth]{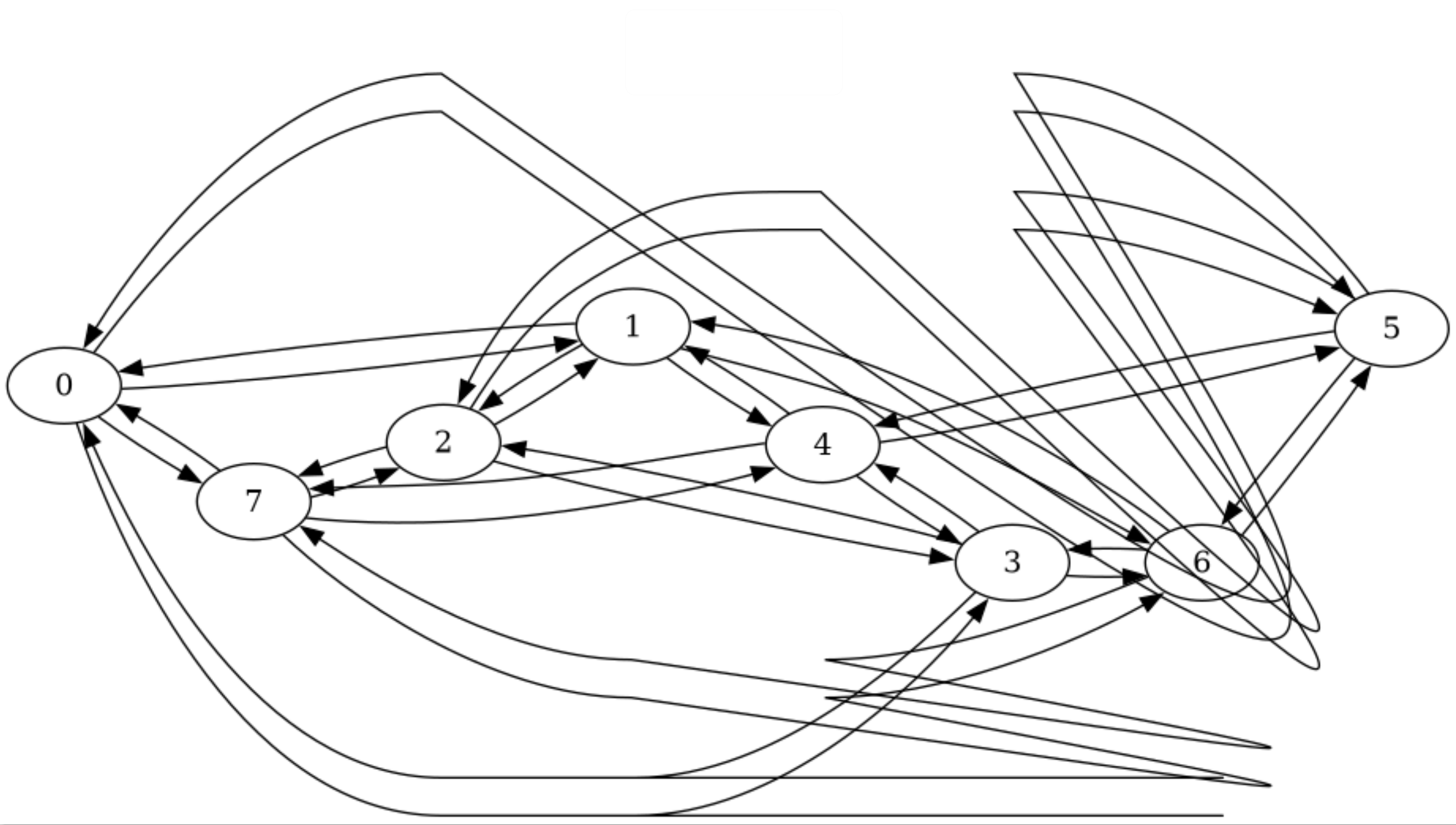}
    \caption{TSS Graph:  Grover [\ref{eq:gr4}]$\;\otimes\; P_x$  }
    \label{fig:gr4_px}
\end{figure}

TSS graph [\ref{fig:gr4_px}] is fully connected and has bidirectional connectivity to every node which implies, every state is reachable from every other state. The basic Grover state is also fully connected, but the newer gate seems to not share the self loop structure. \\

\noindent
$\mathbf{P_z \otimes \mathrm{\textbf{Grover}}}$

We get a matrix whose dimensions are $2^3\times 2^3$ with the highest state vector represented in quantum state notation $\ket{111} \sim \ket{7}$. \\

\begin{figure}[h]
    \centering
    \includegraphics[width=1\linewidth]{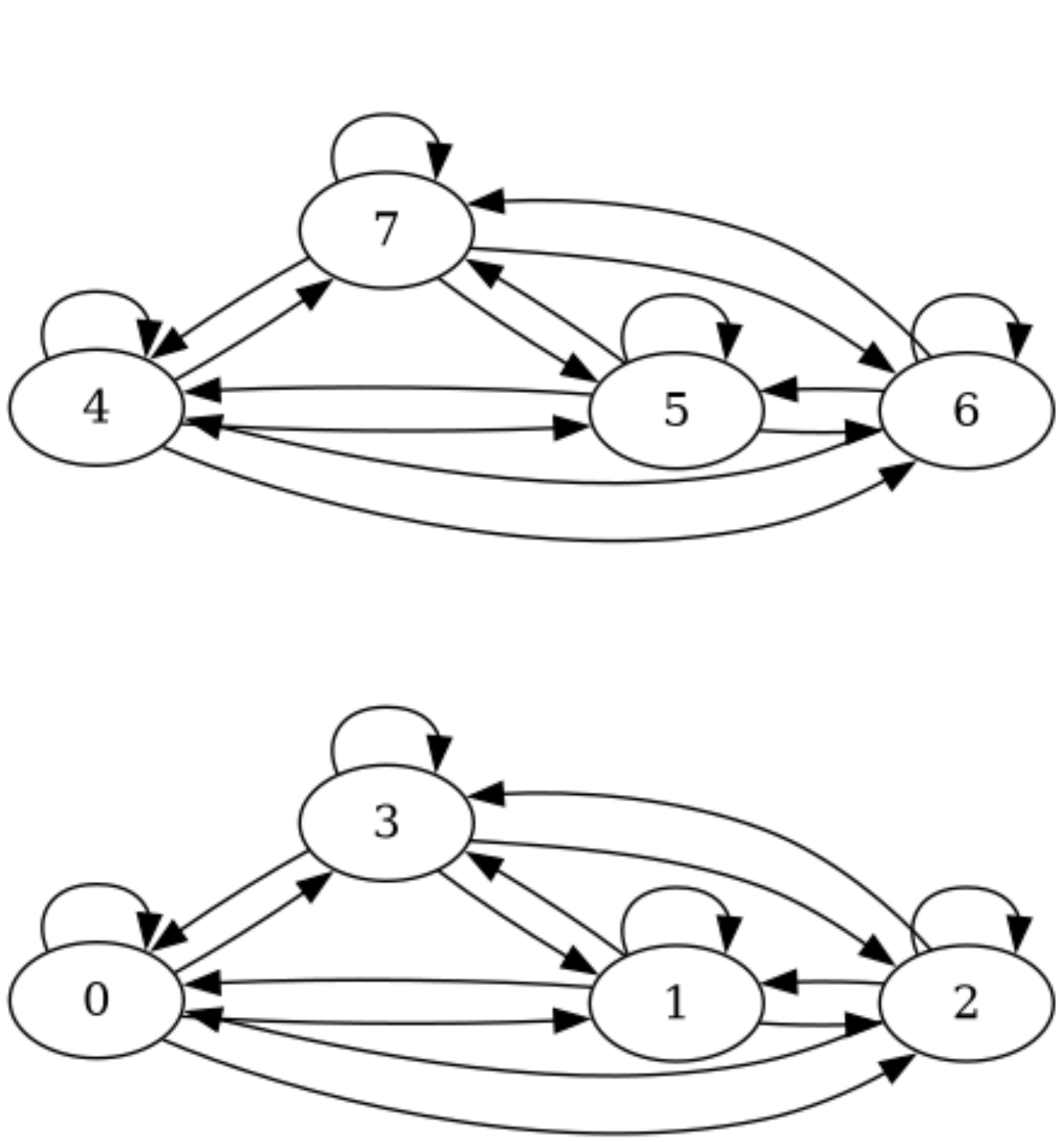}
    \caption{TSS Graph:  $P_z \otimes$ Grover [\ref{eq:gr4}]}
    \label{fig:pz_gr4}
\end{figure}

TSS graph [\ref{fig:pz_gr4}] is broken into two island graphs and self loops exist which seems to have been inherited from the basic Grover state. \\

\noindent
\textbf{Properties of TSS Graphs}
\begin{table}
    \centering
    \begin{tabular}{|l|l|c|c|c|c|c|l|}\hline
          name&bb [\ref{sec:bb}] 
              &saneg [\ref{sec:saneg12}]
              &gr4 [\ref{sec:gr4}]
              &had16 [\ref{sec:had16}]
              &pxpxpxpx [\ref{sec:pxpxpxpx}] \\ \hline
 no. of sinks & 16 & 4 & 4 & 16 & 16 \\ \hline
 no. of sources & 16 & 4 & 4 & 16 & 16 \\ \hline
 sink to source ratio & 1 & 1 & 1 & 1 & 1 \\ \hline
          no. of self loops & 16 & 4 & 4 & 16 & 0 \\ \hline
          no. of loops & 96 & 5 & 24 & 1M+ & 8 \\ \hline
          multiplicity & 4 & 1.5 & 1.5 & 16 & 1 \\ \hline
    \end{tabular}
    \label{tab:placeholder}
\end{table}

\begin{itemize}
    \item \textbf{Sink}, refers to the node where the arrow head ends in an output state vector
    
    \item \textbf{Source}, refers to the node where the arrow starts from an input state vector 

    \item \textbf{Self Loop}, where the source node points to the same sink, and is the smallest possible loop in a TSS graph 

    \item \textbf{Loop}, where directed arrows start and end at the same node covering more than one node

    \item \textbf{Multiplicity}, the ratio of number of inward pointing arrows to a node to that of the no of sources averaged over the entire TSS graph 
\end{itemize}

\begin{figure}[h]
    \centering
    \includegraphics[width=0.8\linewidth]{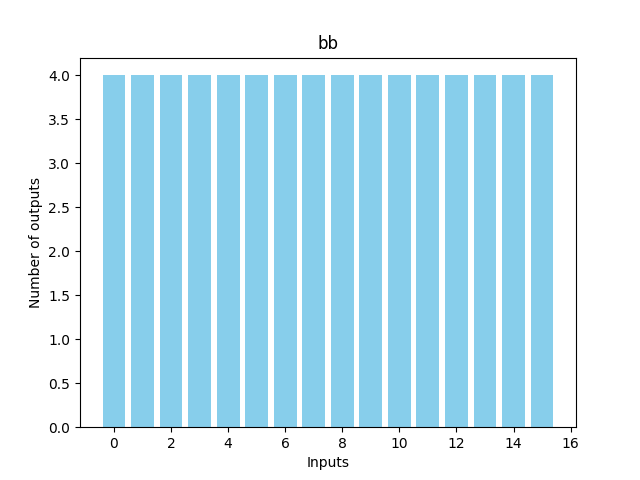}
    \caption{Multiplicity histogram for tensor product of Berkeley $\otimes$ Berkeley [\ref{sec:bb},\ref{fig:bb_full},\ref{fig:bb_node}]}
    \label{fig:bb_full_hist}
\end{figure}

The histograms[\ref{fig:bb_full_hist}] provide insights into how the property of multiplicity helps understand TSS graph structure. Along the X-axis we have the input states, and Y-axis we have the no of times the outputs occurred for that particular input aggregated over all the outputs i.e for the entire TSS graphs. If the shape is a flat square, it indicates all output states are occurring same no of times, but if they are not, then the quantum gate happens to produce only some of the states. Histograms don't reveal connectivity patterns in TSS, for which we to devise another measure. \\

\begin{figure}[h]
    \centering
    \includegraphics[width=0.8\linewidth]{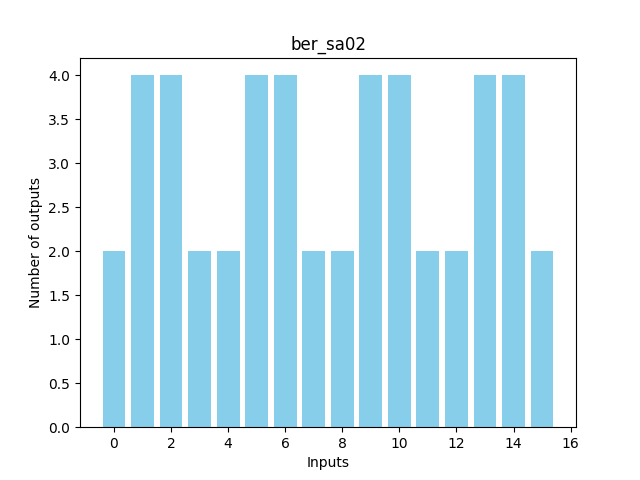}
    \caption{Multiplicity histogram for Berkeley $\otimes$ Swap Alpha 1/2}
    \label{fig:berkeley_swap_alpha_histogram}
\end{figure}

The number of measures we've tabulated above can't be limited. As we get more matrices, and interesting TSS graphs, we have to come up with other measures for analyzing richer substructures in these graphs. These measures help understand the structure of the graph, potentially infer the kind of matrix and hence get a sense of their usefulness in devising Quantum Algorithms. One can notice that TSS graphs have the potential to understand how the matrix should be modified towards an algorithm. \\

\section{Conclusion}
In this paper, we introduced a new measure called as \textbf{Topological Structure of Superpositions (TSS)}—colloquially termed "Imperfect Graphs"—as a novel topological analysis approach for unitary operators. We call the graphs \emph{imperfect} because of lack of any kind of appealing global property. By mapping basis states to vertices and non-zero transitions to edges, we constructed a discrete dynamical view of quantum circuits \cite{childs2009,aharonov2001} that isolates structural connectivity from probabilistic amplitude. TSS graph contains information about all the qubit combinatorics for that particular operator providing a global transition path graph. \\

Our framework reveals a fundamental topological dichotomy between classical-reversible operations and quantum-interference operations. We demonstrated that classical gates manifest as sparse, 1-regular permutation graphs, whereas key quantum subroutines, such as the Hadamard manifest as dense, highly connected graphs. This confirms that the computational power of quantum algorithms is topologically linked to the density of state-space connectivity \cite{aaronson2013}.\\

This visualization technique validates the Unitary Matrix Approach (UMA) by providing a tractable method for analyzing global operator structure \cite{shende2006,vatan2004}. The TSS framework complements existing methods for quantum circuit analysis, including stabilizer formalism \cite{gottesman1997,aaronson2004} and graph state representations \cite{hein2004,vandennest2004,raussendorf2001}, by offering a purely topological perspective. \\

Future work will extend this framework to include probability and phase analysis, potentially connecting TSS topology with known complexity classes \cite{aaronson2013,bravyi2016}. Additionally, we plan to investigate whether TSS patterns can inform automated circuit synthesis \cite{dawson2006} and provide new insights into quantum algorithm design. \\ 

On a final note it should be understood that the TSS method originated to describe Unitary Matrices. It can however be applied to matrices of any kind to understand the properties as Mathematical abstractions for constructing homomorphisms between \emph{matrices} and \emph{graphs}. \\

\printbibliography
\appendix
\section{Unitary Matrices}

In this appendix, we are describing the basic gates used to construct more sophisticated gates that expand the dimensionality and more importantly understanding how these operations can inherit the properties for a richer TSS graph. \\

\noindent
\textbf{Pauli Matrices}\\
Pauli Matrices were devised by Wolfgang Pauli~\cite{pauli1940connection} to theoretically represent the effects of quantization of magnetic interactions. It was later extended to other areas, and generalizations were attempted. One tends to study them in the context of Super Unitary group, SU(n) as well \\

The basic Pauli Matrices have beautiful properties when operated on other matrices especially relevant to Quantum Computing operators. \\

\begin{equation*}
P_x = 
\begin{bmatrix} 0 &1 \\ 1 &0 \end{bmatrix},\quad
P_y = 
\begin{bmatrix} 1 &-i \\ i &0 \end{bmatrix},\quad
P_z = 
\begin{bmatrix} 1 &0 \\ 0 &-1 \end{bmatrix}
\end{equation*}\\

\noindent
\textbf{Hadamard Matrices}\label{sec:had16}\\
They were devised~\cite{hadamard1893resolution} during the late 1800s and used in a wide variety of areas in both science and engineering. Due to their ubiquity, an attempt was made to generalize and deduce all possible matrices. As of today a lot of databases are available with such matrices of increasing orders discovered as more and more computing power~\cite{wallis2006hadamard} is available. \\

On a similar note, a new class of Hadamard matrices called as Complex Hadamard matrices~\cite{tadej2006concise} have been more recently pursued which have odd orders. These are particularly relevant to Quantum Computing unlike their real counterparts which have been used in applications pertaining more to the domain of Signal Processing, Image Processing or the like. \\

For the purpose of this analysis, we've selected one matrix amongst the several matrices documented by Neil Sloane~\cite{sloane1999library}. Note that, this is not a unique Hadamard matrix of order 16, because one can have plenty such of same order. 
\begin{equation*}
    \mathrm{had16}=\left[\begin{smallmatrix}
    1 & 1 & 1 & 1 & 1 & 1 & 1 & 1 & 1 & 1 & 1 & 1 & 1 & 1 & 1 & 1\\
1 & -1 & 1 & -1 & 1 & -1 & 1 & -1 & 1 & -1 & 1 & -1 & 1 & -1 & 1 & -1\\
1 & 1 & -1 & -1 & 1 & 1 & -1 & -1 & 1 & 1 & -1 & -1 & 1 & 1 & -1 & -1\\
1 & -1 & -1 & 1 & 1 & -1 & -1 & 1 & 1 & -1 & -1 & 1 & 1 & -1 & -1 & 1\\
1 & 1 & 1 & 1 & -1 & -1 & -1 & -1 & 1 & 1 & 1 & 1 & -1 & -1 & -1 & -1\\
1 & -1 & 1 & -1 & -1 & 1 & -1 & 1 & 1 & -1 & 1 & -1 & -1 & 1 & -1 & 1\\
1 & 1 & -1 & -1 & -1 & -1 & 1 & 1 & 1 & 1 & -1 & -1 & -1 & -1 & 1 & 1\\
1 & -1 & -1 & 1 & -1 & 1 & 1 & -1 & 1 & -1 & -1 & 1 & -1 & 1 & 1 & -1\\
1 & 1 & 1 & 1 & 1 & 1 & 1 & 1 & -1 & -1 & -1 & -1 & -1 & -1 & -1 & -1\\
1 & -1 & 1 & -1 & 1 & -1 & 1 & -1 & -1 & 1 & -1 & 1 & -1 & 1 & -1 & 1\\
1 & 1 & -1 & -1 & 1 & 1 & -1 & -1 & -1 & -1 & 1 & 1 & -1 & -1 & 1 & 1\\
1 & -1 & -1 & 1 & 1 & -1 & -1 & 1 & -1 & 1 & 1 & -1 & -1 & 1 & 1 & -1\\
1 & 1 & 1 & 1 & -1 & -1 & -1 & -1 & -1 & -1 & -1 & -1 & 1 & 1 & 1 & 1\\
1 & -1 & 1 & -1 & -1 & 1 & -1 & 1 & -1 & 1 & -1 & 1 & 1 & -1 & 1 & -1\\
1 & 1 & -1 & -1 & -1 & -1 & 1 & 1 & -1 & -1 & 1 & 1 & 1 & 1 & -1 & -1\\
1 & -1 & -1 & 1 & -1 & 1 & 1 & -1 & -1 & 1 & 1 & -1 & 1 & -1 & -1 & 1\\
\end{smallmatrix}\right]
\end{equation*}

\noindent
\textbf{Swap Alpha}\\
This is a two qubit gate~\cite{blaauboer2008analytical} which has a phase parameter inherently attached to the matrix. By modifying the parameters we get quite a variety of gates with interesting properties
\begin{equation*}
    e^{i\pi/2\alpha}\begin{bmatrix}
        e^{-i\pi/2\alpha} &0 &0 &0 \\
        0 &\cos(\pi/2\alpha) &i\sin(\pi/2\alpha) &0 \\
        0 &i\sin(\pi/2\alpha) &\cos(\pi/2\alpha) &0 \\
        0 &0 &0 &e^{-i\pi/2\alpha} \\
    \end{bmatrix}
\end{equation*}\\

\noindent
\textbf{Berkeley Gate}\\
B-Gate~\cite{zhang2004minimum} as it is often referred to is a two qubit gate that shares its place as an important gates whose applications are yet to be discovered. 

\begin{equation*}
    b=\begin{bmatrix}
        \cos(a) &0 &0 &i\sin(a) \\
        0 &\cos(b) &i\sin(b) &0 \\
        0 &i\sin(b) &\cos(b) &0 \\
        i\sin(a) &0 &0 &\cos(a)\\
    \end{bmatrix},\;a=\pi/8;\quad b=3\pi/8
\end{equation*}\\

\noindent
\textbf{Swap Alpha(-1/2)}\label{sec:saneg12}\\
\begin{equation*}
    \mathrm{saneg12}=\begin{bmatrix}
    1 & 0 & 0 & 0\\
0 & 0.5 - 0.5i & 0.5 + 0.5i & 0\\
0 & 0.5 + 0.5i & 0.5 - 0.5i & 0\\
0 & 0 & 0 & 1\\
\end{bmatrix}
\end{equation*}

\noindent
\textbf{Swap Alpha(1/2)}\label{sec:sapos12}\\
This is the square root of SWAP alpha gate, where $\alpha=\pi/4$
\begin{equation*}
    \mathrm{sapos12}=\begin{bmatrix}
    1 & 0 & 0 & 0\\
    0 & 0.5 + 0.5i & 0.5 - 0.5i & 0\\
    0 & 0.5 - 0.5i & 0.5 + 0.5i & 0\\
    0 & 0 & 0 & 1\\
\end{bmatrix}
\end{equation*}

\noindent
\textbf{Grover 2-qubit Gate}\label{sec:gr4}\\
This is a $2$-qubit Grover's gate constructed just to check how TSS looks like. Grover's algorithm~\cite{grover1996} is a fundamental algorithm and possibly the only algorithm that modifies the probability factors iteratively on a given superposed state without modifying the superposed states. \\

The generalization of this algorithm involves constructing a Quantum Gate that doesn't parameterize the gate to contain the state being searched for. Constructing an operator agnostic Gate involves more sophisticated heuristics where feedback loops might be necessary. Such a Quantum Operator as of writing this paper hasn't been devised yet. 
\begin{equation*}
gr4=\begin{bmatrix}
    -0.5 & 0.5 & 0.5 & 0.5\\
0.5 & -0.5 & 0.5 & 0.5\\
0.5 & 0.5 & -0.5 & 0.5\\
0.5 & 0.5 & 0.5 & -0.5\\
\end{bmatrix}
\label{eq:gr4}
\end{equation*}

\section{TSS Graphs}

The word "full" in a figure caption indicates that all the states are combined to form the \emph{complete TSS graph} as opposed to \emph{node level TSS graph}. In some cases both types are similar because there's not much connectivity amongst the individual nodes. \\

\begin{figure}[h]
    \centering
    \includegraphics[width=0.5\linewidth]{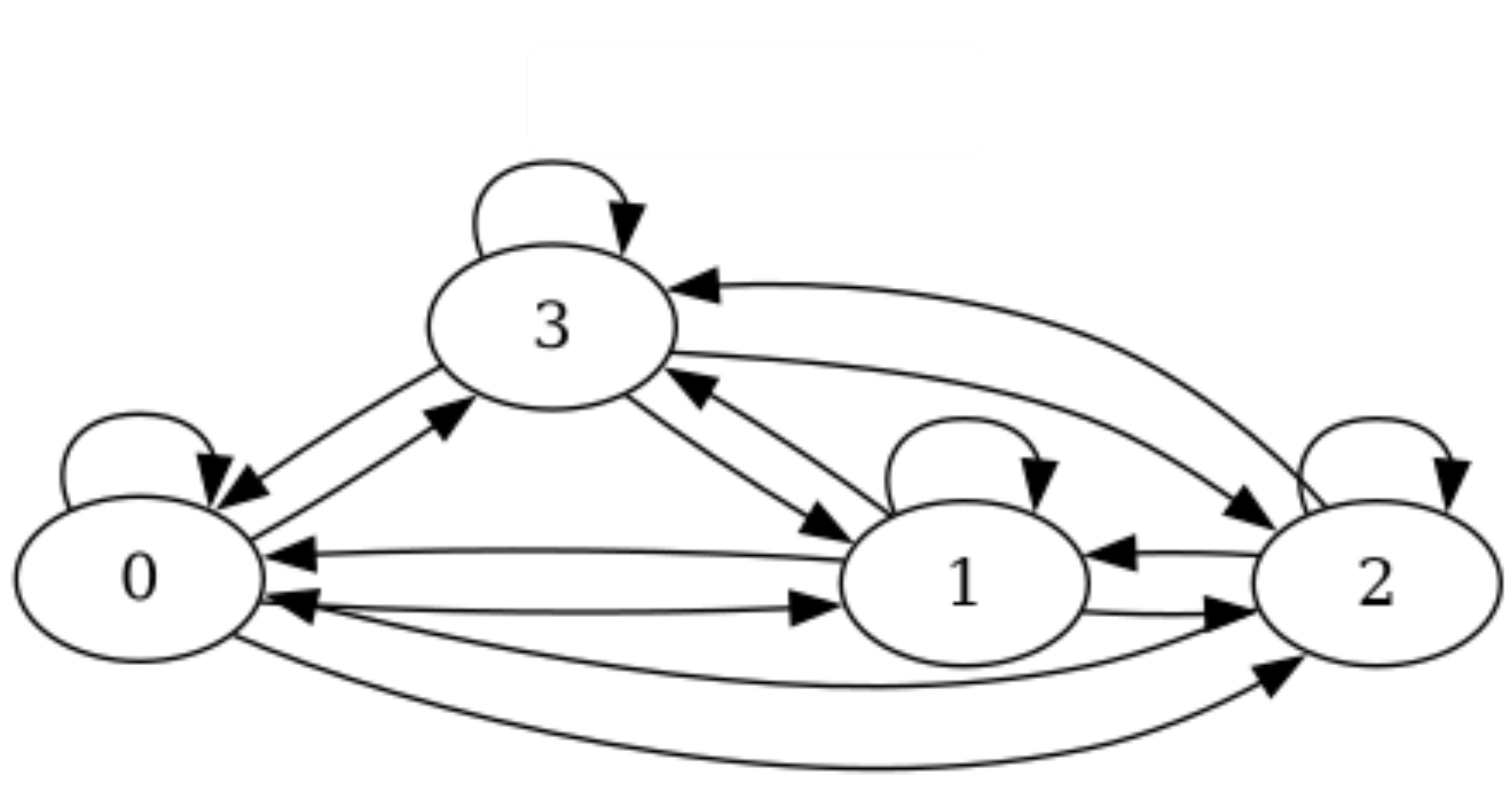}
    \caption{TSS Graph: Grover 2-qubit [\ref{eq:gr4}]}
    \label{fig:gr4_full}
\end{figure}

\begin{figure}[h]
    \centering
    \includegraphics[width=1\linewidth]{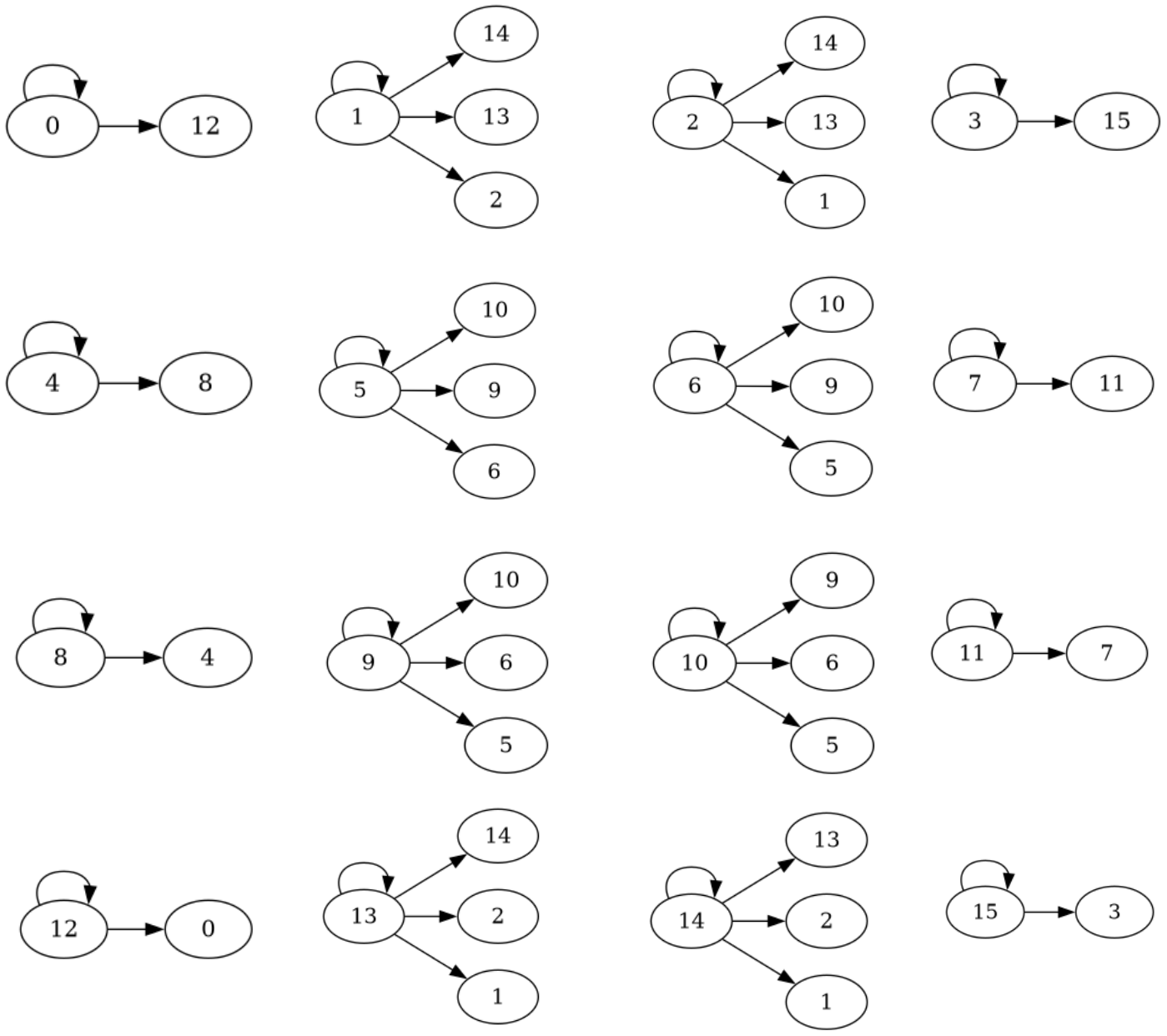}
    \caption{Node Level TSS: Berkeley $\otimes$ Swap Alpha(1/2)}
    \label{fig:ber_sa001}
\end{figure}

\begin{figure}[h]
    \centering
    \includegraphics[width=0.5\linewidth]{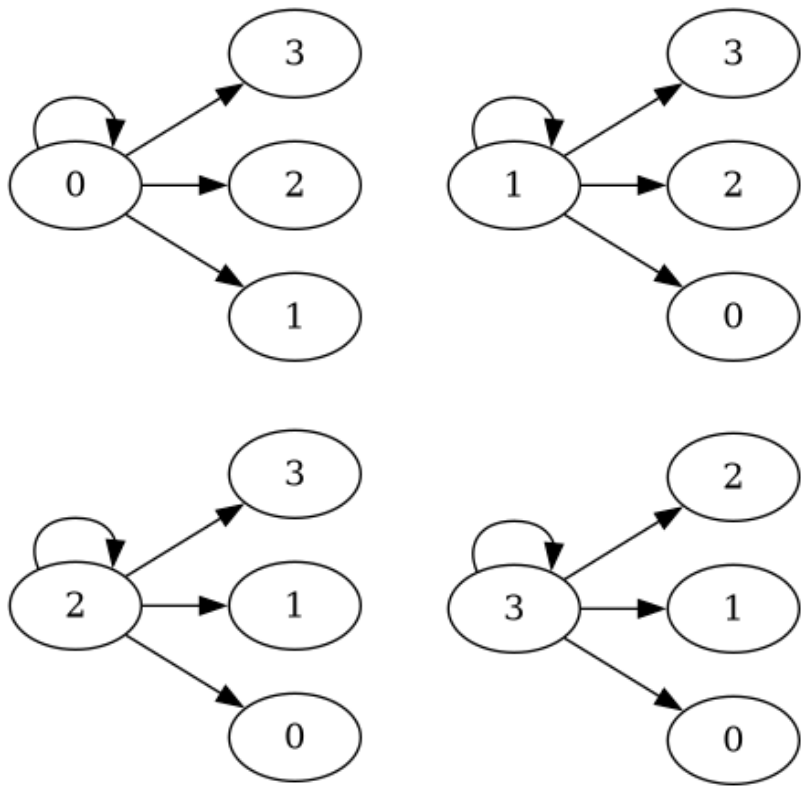}
    \caption{Node Level TSS: Grover [\ref{eq:gr4}]}
    \label{fig:gr4_node}
\end{figure}

\begin{figure}[h]
    \centering
    \includegraphics[width=0.5\linewidth]{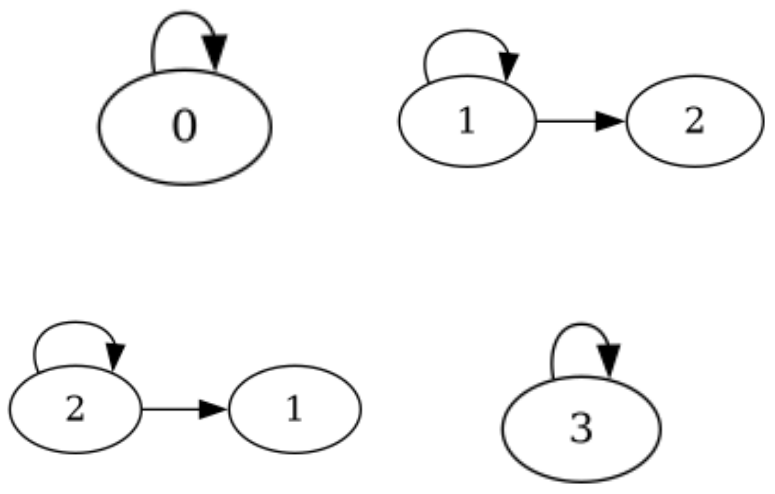}
    \caption{TSS Graph Swap Alpha(-1/2)}
    \label{fig:placeholder}
\end{figure}

\end{document}